\documentclass[usenatbib,useAMS]{mnras}
\usepackage{mystyle}

\title[]{Theories with higher-order time derivatives and the Ostrogradsky ghost}

\author[Svanberg]{
Eleonora Svanberg\thanks{E-mail: \href{mailto:eleonora.svanberg@fysik.su.se}{eleonora.svanberg@fysik.su.se}}\\
}
\begin{document}
\maketitle

\begin{wrapfigure}[9]{r}{0.35\textwidth}
    \centering
    \includegraphics[width=0.35\textwidth]{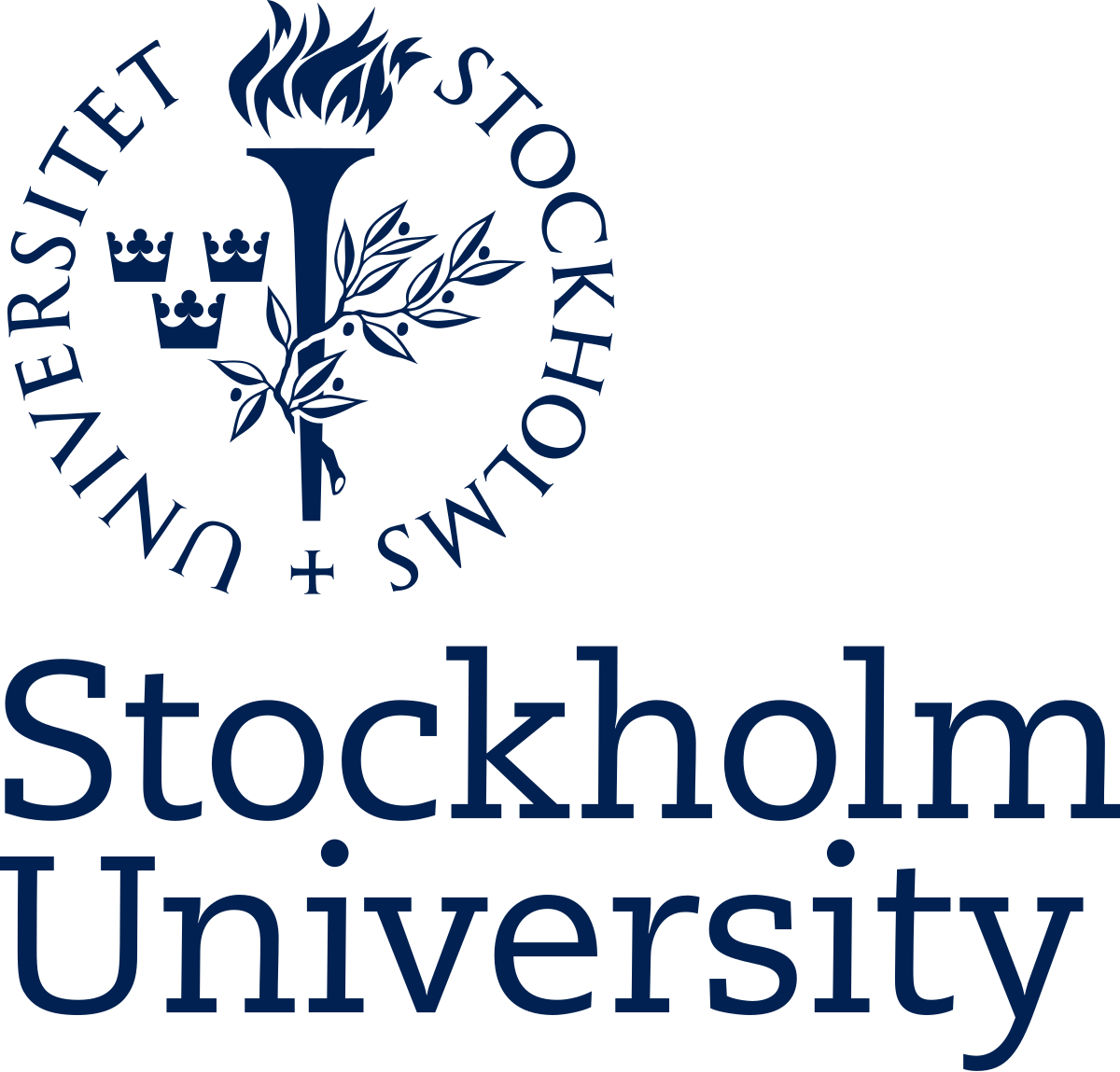}
\end{wrapfigure}
\vspace{20pt}
\hfill \\
\hfill

\noindent
{\Large Supervisors: Dr. Fawad Hassan, Joakim Flinckman}\\
Stockholms Universitet/Stockholm University \\
Department of Physics \\
Bachelor's Degree \\
Physics, Bachelor Project, 15 ECTS 2022 \\
SE-106 91 Stockholm\\
\href{www.su.se}{www.su.se}\\

%%%%%%%%%%%%%%%  ABSTRACT %%%%%%%%%%%%%%%%%%%%%%%%%%%%%%%%%%
\newpage
\begin{abstract}
\begin{center}
    \justify{
Newton's second law, Schrödinger's equation and Maxwell's equations are all theories\\ composed of at most second-time derivatives. Indeed, it is not often we need to take the time derivative of the acceleration. So why are we not seeing more higher-order derivative theories? Although several studies present higher derivatives' usefulness in quadratic\\ gravity and scalar-field theories, one will eventually encounter a problem. In 1850, the physicist Mikhail Ostrogradsky presented a theorem that stated that a non-degenerate Lagrangian composed of finite higher-order time derivatives results in a Hamiltonian unbounded from below. Explicitly, it was shown that the Hamiltonian of such a system includes linearity in physical momenta, often referred to as the ''Ostrogradsky ghost''. This thesis studies how one can avoid the Ostrogradsky ghost by considering degenerate Lagrangians to put\\ constraints on the momenta. The study begins by showing the existence of the ghost and later cover the essential Hamiltonian formalism needed to conduct Hamiltonian constraint analyses of second-order time derivative systems, both single-variable and systems coupled to a regular one. 
Ultimately, the degenerate second-order Lagrangians successfully eliminate the Ostrogradsky ghost by generating secondary constraints restricting the physical\\ momenta. Moreover, an outline of a Hamiltonian analysis of a general higher-order\\ Lagrangian is presented at the end. }
\end{center}

% Furthermore, the general $n$th-order-derivative energy function is derived from Noether's theorem and used to examine the existence of the Ostrogradsky ghost, something which cannot be found in the literature to the author's knowledge. 
% The Hamiltonian is shown to be linear in the physical momenta for a non-degenerate Lagrangian of finite order $n>1$. 

\end{abstract}

%%%%%%%%%%%%%%%%%%%%%%%%%%%%%%%%%%%%%%%%%%%%%%%%%
\newpage
\tableofcontents
\newpage
%%%%%%%%%%%%%%%%%%%%%%%%%%%%%%%%%%%%%%%%%%%%%%%%%
\pagenumbering{arabic}
\setcounter{page}{3}

\section{Introduction}\label{in}
When introduced to physics, a student's first encounter is most likely the well-known Newton's second law 
\begin{equation}\label{eq:N2}
    m \ddot{x} = F(\dot{x}, x, t),
\end{equation}
connecting the second-order time derivative of the position to the force \citep{newton}. Shortly after, the student will learn about more renowned equations such as Maxwell's and  Schrödinger's equations. Both of which includes the first-order time derivative. A natural question, but not as common as one might think, is to wonder \textit{why} these theories are of at most second-order time derivatives. Why not say the third? Fourth? Or nth? This question puzzled the physicists who started to research higher derivative theories.

It turns out there are several benefits when considering a system with no restriction on the order of derivatives. Its application can be found in several studies on higher derivative theories, including non-local field theory \citep{Erbin_2022} and general higher-order scalar-tensor (Horndeski) theories \citep{Gleyzes_2015,Zumalac_rregui_2014, Lin_2014}. Moreover, infinite derivative theories of general relativity, or quadratic gravity, have recently been found to be renormalisable \citep{Salvio_2018}.

Undoubtedly, describing a system using a higher-order derivative Lagrangian is of interest. However, in 1850, Ostrogradsky presented the Ostrogradsky Theorem, which showed that a finite higher-order non-degenerate Lagrangian leads to an unbounded Hamiltonian \citep{ostrogradsky}. We will in section \ref{sec:introham} explore what degeneracy means in the Lagrangian context. Although for now, it is enough to think of a non-degenerate Lagrangian as a Lagrangian composed of independent variables. Anyhow, Ostrogradsky demonstrated that these theories include a ''ghost'', meaning there is an unphysical variable in the theory affecting the physical property of the system. The ghost appears in the form of negative kinetic energy, linear physical momentum in the Hamiltonian and unstable degrees of freedom. As a result, the Hamiltonian is unbounded from below and a regular one-derivative system would be able to absorb infinite energy if coupled to an Ostrogradsky system. In other words, the Ostrogradsky ghost (or sometimes instability) makes the theory fundamentally ''sick'' and one wants to avoid them to construct a physical theory. ''Healthy'', or ghost-free, higher-order theories have thus been a topic of interest for physicists for a long time. The main focus of this thesis will be on the context of the Ostrogradsky instability; can we use higher-order derivatives and ensure a physical Hamiltonian, free from the Ostrogradsky ghost? In fact, it was recently shown that degenerate second-order Lagrangians generate constraints that can successfully evade the Ostrogradsky ghost \citep{Ganz}. This thesis will explore the concepts around the Ostrogradsky ghost and, more importantly, how to avoid it. The focus will be on degenerate systems and the Hamiltonian formalism, mainly primary and secondary constraints, the kinetic matrix and pursuing a Hamiltonian constraint analysis. 
%Additionally, we will present the energy function for an arbitrary higher-order derivative Lagrangian and examine the existence of the Ostrogradsky ghost, which to the author's knowledge cannot be found in the literature.

The thesis is organised as follows. First, in section \ref{sec:existence}, we will show the existence of the Ostrogradsky ghost from kinetic energy and momenta, and then introduce Lagrange multipliers to allow us to apply the usual Hamiltonian formalism later. After that, in section \ref{sec:introham}, we will go through the essentials of Hamiltonian formalism, including primary and secondary constraints and the Dirac algorithm. Then, in section \ref{sec:hamanalysis} we will conduct a Hamiltonian constraint analysis of second-order Lagrangian, both single-variable and coupling with a regular system. An outline for future Hamiltonian analysis of a system with general order derivatives is later presented in section \ref{sec:genHA}. Finally, the main results are discussed in section \ref{sec:cd} 

%The general energy function for a $n$th-order Lagrangian is derived in section \ref{sec:nthenergy} and discussed in \ref{sec:efunc}.

%%%%%%%%%%%%%%%%% 2. %%%%%%%%%%%%%%%%%%%%% 
\section{The existence of the Ostrogradsky ghost}\label{sec:existence}
Before considering the general case of higher-order Lagrangians, let us consider perhaps the most trivial example and see how the Ostrogradsky ghost can present itself in different ways. This will also serve as an example of how different Lagrangians can provide equivalent equations of motion.

The section will explore how the Ostrogradsky ghost leads to negative kinetic energy and the Ostrogradsky signature; a Hamiltonian that is linear in physical momenta. The systems we will be considering in this section will initially all be of independent variables, i.e. the initial Lagrangians will be non-degenerate.

%%%%%%%%%%%%%%%%% Example %%%%%%%%%%%%%%%%%%%%% 
We begin by considering the simple second-order Lagrangian of the coordinate $\phi(t)$. In this thesis, $\phi$ will always be used as a coordinate and should not be confused with a field. The action of the system is given by 
\begin{equation}\label{eq:example}
    L(\phiddot, \phidot,\phi) = \tfrac{1}{2}\phiddot^2, \quad S[\phi] = \int \text{d}t \, \tfrac{1}{2}\phiddot^2,
\end{equation}
where we have used the convention $\phidot=\mathrm{d}\phi/\mathrm{d}t$.
Of course, the Euler Lagrange (EL) equation, or Equations of Motion (EoM) will look different for a Lagrangian of second-order derivatives. However, the derivation is similar to the usual case. The EL equation for a general second-order Lagrangian of a variable $\phi$, thus reads 
\begin{equation}
        \pL{\phi} - \frac{\text{d}}{\text{d}t}\left(\pL{\phidot}\right)+\frac{\text{d}^2}{\text{d}t^2}\left(\pL{\phiddot}\right) %= \pL{\phi} - \dt \left\{\left(\pL{\phidot}\right)- \dt\left(\pL{\phiddot}\right)\right\} 
        = 0,
\end{equation}
and is derived using variational principle, see Appendix \ref{sec:EL} for full derivation. For this example, the equation of motion simply reads
\begin{equation}\label{eq:ELexmp}
    \phi: \quad \ddot{\phiddot} = 0,
\end{equation}
which is largely different to Newton's second law and Maxwell's equations.
There are plenty of methods to show that a Lagrangian composed of a \textit{exotic} variable (their derivatives appear more than to first order, in comparison to a \textit{regular} variable) will always generate an unbounded Hamiltonian, causing problem when the system is coupled to a regular system. he following subsections will show the existence of an Ostrogradsky ghost, by two different methods.

\subsection{Negative kinetic energy}
In order to get a clear view of the ghost, we rewrite the Lagrangian into one with equivalent EoM (hence representing the same system) by introducing a new auxiliary variable $\psi(t)$ as $\psi := \phiddot$. Using the fact that
\begin{equation}
    \tfrac{1}{2}(\phiddot-\psi)^2 = \tfrac{1}{2}\phiddot^2+\tfrac{1}{2}\psi^2 - \phiddot\psi = 0,
\end{equation}
and integration by parts, we can rewrite the action \eqref{eq:example} as
\begin{align}
    S[\phi, \psi] &=   \int \text{d}t \, \left( \phiddot \psi-\tfrac{1}{2} \psi^{2}\right)  = \int \text{d}t \, \left(-\phidot \dot{\psi}-\tfrac{1}{2} \psi^{2}\right) +\left[\dot{\phi}\psi \right],
\end{align}
where the last term is a boundary term (BT). A boundary term in the action does not affect the equations of motion as they are derived by the principle of least action —— and a BT will not affect the outcome. We can thus only consider the expression inside the integral, without losing information about the system. The new, equivalent, Lagrangian is now 
\begin{equation}
    \Tilde{L}(\phidot, \phi, \dot{\psi}, \psi ) =-\phidot\dot{\psi}-\frac{1}{2}\psi^2,
\end{equation}
composed of two regular variables. The EL equations for regular variables read
\begin{subequations}
\begin{align}
&\phi: \quad \frac{\p \Tilde{L}}{\p \phi}- \dt \left(\frac{\p \Tilde{L}}{\p \phidot} \right) = \ddot{\psi}=0, \\
&\psi: \quad \frac{\p \Tilde{L}}{\p \psi}- \dt \left(\frac{\p \Tilde{L}}{\p \dot{\psi}} \right) = \psi - \phiddot = 0, 
\end{align}
\end{subequations}
which immediately yields $\ddot{\psi}=\ddot{\phiddot}=0$, and we have confirmed that they have the same dynamical equations and thus describe the same system. We will now diagonalise these variables into two new ones to separate into one "healthy" and one "ghost" variable. Consider the variable change
\begin{equation}
    \Phi = \frac{\phi -\psi}{\sqrt{2}}, \quad \Psi= \frac{\phi +\psi}{\sqrt{2}},
\end{equation}
and the equivalent Lagrangian 
\begin{equation}
    \widetilde{L}(\dot{\Phi}, \Phi, \dot{\Psi}, \Psi) = \tfrac{1}{2}\dot{\Phi}^2-\tfrac{1}{2}\dot{\Psi}^2-\tfrac{1}{4}(\Phi-\Psi)^2.
\end{equation}
We can see that the $\Psi$ is a ghost variable as it appears in the Lagrangian with a negative kinetic term, a signalment of a ghost. However, let us see what the Hamiltonian looks like for this system for redundancy. The usual definition of canonical momenta gives
\begin{equation}
    P_\Phi = \frac{\p  \widetilde{L}}{\p\dot{\Phi}}=\dot{\Phi}, \quad P_\Psi = \frac{\p  \widetilde{L}}{\p\dot{\Psi}}=-\dot{\Psi},
\end{equation}
and using this in the Legendre transformation, the final Hamiltonian becomes
\begin{equation}
\begin{split}
    H &= P_{\Phi}^2 - P_{\Psi}^2 - \left[ \tfrac{1}{2}P_{\Phi}^2 - \tfrac{1}{2}P_{\Psi}^2-\tfrac{1}{4}(\Phi-\Psi)^2\right], \\
    &= \tfrac{1}{2}P_{\Phi}^2-\tfrac{1}{2}P_{\Psi}^2+\tfrac{1}{4}(\Phi-\Psi)^2.
\end{split}
\end{equation}
Evidently, we get a negative kinetic term of $\Psi$ and a clear example of the Ostrogradsky ghost. Although the total energy of the system will be conserved, since we have a $\Phi\Psi$-term, $\Phi$ and $\Psi$ will interact. The negative kinetic term will result in that $\Phi$ would therefore be able to absorb infinite energy from $\Psi$, something that is clearly unphysical.

\subsection{Linear physical momenta}\label{sec:physicalmomenta}
In general, it will not be possible to diagonalise variables and separate the ghost explicitly, so we will now use a different method to approach the ghost, using the same example. This section aims to show the physical momenta's role in the Ostrogradsky context. The details behind this approach require concepts not yet discussed, so the reader is advised to take the following arguments lightly and read section \ref{sec:hamanalysis} for an in-depth analysis, however, this section will give the overview needed to get its gist. We will now start with finding the expression for the physical momenta, then introducing Lagrange multipliers and finally end with applying the knowledge to the example \eqref{eq:example}.

We cannot assume that the physical momenta takes the standard form of $\pdv{L}{\dot{\phi}}$, and we will see that it is in fact modified with the presence of higher-derivative terms. By definition, momenta is what is preserved when the Lagrangian is coordinate invariant. In our case, we can see that if $L$ is $\phi$ invariant, then the derivative with respect to $\phi$ in the EL equation \eqref{eq:ELexmp} is zero. That means that the expression inside the curly brackets is the physical momenta as that what's conserved;
\begin{equation}\label{eq:physical momentum}
    \pL{\phi} = \dt \left\{\left(\pL{\phidot}\right)- \dt\left(\pL{\phiddot}\right)\right\} = \dt \left\{\text{physical momentum} \right\}.
\end{equation}

Before we get back to the example, we need to sort out how the expression of physical momenta appears when we make a change of variables. Consider a general second-order Lagrangian $L(\phiddot, \phidot, \phi)$ of one variable. In order to find an equivalent system with regular variables, we introduce a Lagrange multiplier $\lambda$ as a new coordinate. Now setting $q:=\phidot$ the joint action of the system can be written like
\begin{equation}
    S[q, \phi,\lambda] = \int \text{d}t \, \left( L(\qdot, q, \phi) + \lambda(\phidot-q)\right) = \int \text{d}t \, L_{\text{tot}}(\qdot, q, \phidot, \phi, \lambda).
\end{equation}
The exotic single-variable system has now been reduced to a system with three regular variables $q, \phi$ and $\lambda$. That this system is identical to what is presented in \eqref{eq:example} can be shown by the EL equations for $\phi, q$ and $\lambda$:
\begin{subequations}
\begin{align}
    &\phi: \quad \frac{\p L_{\text{tot}}}{\p \phi} - \dt \left( \frac{\p L_{\text{tot}}}{\p \phidot}\right) = \pL{\phi} -\dot{\lambda} = 0,\\ 
    &q: \quad \frac{\p L_{\text{tot}}}{\p q} - \dt \left( \frac{\p L_{\text{tot}}}{\p \qdot}\right) =\pL{q} - \lambda -\dt \left( \pL{\qdot}\right)=0, \label{eq:star1} \\ 
    &\lambda: \quad \frac{\p L_{\text{tot}}}{\p \lambda} - \dt \left( \frac{\p L_{\text{tot}}}{\p \dot{\lambda}}\right) =\phidot-q = 0, \label{eq:star2}
\end{align}
where we can see that the EoM for $\lambda$ \eqref{eq:star2} exactly yields our definition $q=\phidot$.
\end{subequations}
Using this and taking the derivative of the $q$ equation \eqref{eq:star1} and inserting $\dot{\lambda}=\tfrac{\p L}{\p \phi}$ gives
\begin{equation}
    \dt \left( \pL{q} \right) - \pL{\phi} - \ddtt \left( \pL{\qdot} \right) = 0 \;\;\; \Longrightarrow       \;\;\;   \pL{\phi} - \frac{\text{d}}{\text{d}t}\left(\pL{\phidot}\right)+\frac{\text{d}^2}{\text{d}t^2}\left(\pL{\phiddot}\right)=0.
\end{equation}
We have thus shown that this is an equivalent system to a general second-order Lagrangian system, and can now continue with a Hamiltonian analysis. Now, using the definition of canonical momenta, we obtain
\begin{align}
    P_\phi = \frac{L_{\text{tot}}}{\p \phidot} =\lambda, \quad   p = \frac{L_{\text{tot}}}{\p \qdot} =\pL{\qdot}, \quad P_\lambda = \frac{L_{\text{tot}}}{\p \dot{\lambda}}. 
\end{align}
The physical momenta in the general second-order system is given by the expression \eqref{eq:physical momentum}, can we connect that to the canonical momenta in this system? Using the EL equations and that $P_\phi=\lambda$ reads
\begin{equation}
    P_\phi = \lambda = \pL{q} -\dt \left( \pL{\qdot}\right) = \pL{\phidot}-\dt \left( \pL{\phiddot}\right), 
\end{equation}
and we have obtained the expression given by \eqref{eq:physical momentum}, and thus $P_\phi$ will be the physical momenta of the second-order Lagrangian system $L(\phiddot, \phiddot, \phi)$. The final Hamiltonian of this system is then given by the Legendre transformation
\begin{equation}\label{eq:Hamexp}
\begin{split}
    H &= P_\phi \phidot + p \qdot + P_\lambda \dot{\lambda} - \left[\tfrac{1}{2}p^2 + P_\phi (\phidot - q)\right] \\
    &= P_\phi q+ \tfrac{1}{2}p^2.
\end{split}
\end{equation}
This expression is linear in the physical momenta $P_\phi$, which can take negative values. In other words, that the Hamiltonian is linear in physical momentum means that it is unbounded from below. We have thus shown that the general second-order suffers from the Ostrogradsky ghost.

Finally, let's apply this to our example and set $L$ to \eqref{eq:example}. Again, $q:=\phidot$ and $\lambda$ is the Lagrange multiplier. The joint Lagrangian is then 
\begin{equation}
  L_{\lambda}(\qdot, q, \phidot, \phi, \lambda) = \tfrac{1}{2}\qdot^2 + \lambda(\phidot-q). %\quad \#\dof=6,
\end{equation}
The EL equations are
\begin{subequations}
\begin{align}
&\phi: \quad \frac{\p L_\lambda}{\p \phi}- \dt \left( \frac{\p L_\lambda}{\p \phidot}\right) = -\dot{\lambda} = 0 \label{eq:lmbdeq} \\
&q: \quad \frac{\p L_\lambda}{\p q}- \dt \left( \frac{\p L_\lambda}{\p \qdot}\right) = - (\lambda + \qddot)=0 \label{eq:qelx}\\ 
&\lambda: \quad \frac{\p L_\lambda}{\p \lambda}- \dt \left( \frac{\p L_\lambda}{\p \dot{\lambda}}\right)= (\phidot-q)=0.\label{eq:lamelx}
\end{align}
\end{subequations}
Taking the time derivative of the $q$ equation \eqref{eq:qelx} reads $\dot{\qddot} = -\ddot{\lambda}\lref{\ref{eq:lmbdeq}} 0 = \ddot{\phiddot}$, since $q = \phidot$ from \eqref{eq:lamelx}. We can therefore conclude this is an equivalent system to \eqref{eq:example} as the EL equations are the same. Now, for the canonical momenta: 
\begin{subequations}
\begin{align}
&P_{\phi}=\frac{\p L_\lambda}{\p \dot{Q}_1}=\lambda,\\ 
&p = \frac{\p L_\lambda}{\p \qdot}=\qdot, \\
&P_{\lambda}=\frac{\p L_\lambda}{\p \dot{\lambda}}=0.
\end{align}
\end{subequations}
%We can immediately see that the last equation reduces a degree of freedom. However since the first equation connects $P_1$ and $\lambda$, we have eliminated another degree of freedom. We thus end up with $6-2=4 \, \dof$.
Using standard Legendre transformation, we obtain the same Hamiltonian as in \eqref{eq:Hamexp}. Indeed, the Hamiltonian is unbounded from below. 

How can this be avoided? Going from a system of higher derivatives will not solve the issue, so one can naively want to abandon these theories. However, a solution is found in the loopholes of the Ostrogradsky theorem by considering degenerate systems. One could eliminate the ghost by having a constrained $P_\phi$, but to do so, we require a Hamiltonian analysis with a formalism that is introduced in section \ref{sec:introham}.

\section{Introduction to Hamiltonian formalism}\label{sec:introham}
We have established that the Ostrogradsky instability occurs when we have a non-degenerate Lagrangian of finite higher order derivatives. Additionally, we have connected a non-degenerate Lagrangian to independent variables, which is the case most of us are used to considering. We will in this section sort out what degeneracy is, to later apply it in the Ostrogradsky context. We are now ready for the definition of the Ostrogradsky theorem
\begin{theorem}
A non-degenerate Lagrangian of finite higher-order time derivatives results in a Hamiltonian unbounded from below.
\end{theorem}
There are therefore three cases for which we can avoid the instability; infinite-order derivatives, a degenerate Lagrangian or, of course, having only single time derivatives in the Lagrangian. The aim of this study will be on degenerate Lagrangians and the term ''higher-order derivatives'' will from now on only refer to a finite order.

Indeed, it is necessary to introduce how to deal with constrained momenta. The convention is to do it in the Hamiltonian formalism and work with primary and secondary constraints. Therefore, this section will cover the essential concepts needed to conduct a Hamiltonian analysis of constrained systems.

%%%%%%%%%%%%%%%%% Introduction constraints + degeneracy %%%%%%%%%%%%%%%%%%%%% 
\subsection{Degeneracy}
This section will follow the arguments presented in \citealp{bojowald,Henneaux,Dirac} to introduce degeneracy and constraints, and how they are connected in both Lagrangian and Hamiltonian formalism. We begin by considering a regular Lagrangian\footnote{It is a convention to use superscripts for coordinates and subscripts for canonical momenta. However, it might look confusing at first sight, as it can be confused with the notation for exponents. Hopefully, the context will make it clear. Although a coordinate $q^i$ to the power of $x$ will look like $(q^i)^x$.} $L(\qdot^i,q^i)$ for $i = 1, \dots \mathcal{N}$. From now on, standard Einstein summation convention\footnote{The Einstein summation convention is, in simple words, to assume indexed variables are summed over as $a^ib_i =  \sum_{i=1}^\mathcal{N} a^i b_i $. This notation will make the expressions simpler.} will be used to simplify the expressions.  The Euler-Lagrange equations are
\begin{align}
  0& = \frac{\partial L}{\partial q^i} - \frac{\text{d}}{\text{d}t}\left(\frac{\partial L}{\partial \qdot^i}\right) 
    = \frac{\partial L}{\partial q^i} - \frac{\partial^2L}{\partial \qdot^i \partial q^j }\qdot^j - \frac{\partial^2L}{\partial \qdot^i \partial \qdot^j }\qddot^j.
\end{align}
From this we will now define the \textit{kinetic} matrix $K_{ij}:=\frac{\partial^2L}{\partial \qdot^i \partial \qdot^j}$ and the $M_i$-vector as $M_i:= \frac{\partial L}{\partial q^i} - \frac{\partial^2L}{\partial \qdot^i \partial q^j}\qdot^j$ . The EoM can then be expressed using $K_{ij}$ as
\begin{subequations}
\begin{equation}\label{eq:defK}
    K_{ij}\qddot^j =   M_i(\qdot^i, q^i,) \quad \text{or}   \quad  \textbf{K}\ddot{\textbf{q}} = \textbf{M}(\dot{\textbf{q}}, \textbf{q}),
\end{equation}
or in matrix form 
for $\textbf{q}:=(q^1, \dots, q^\mathcal{N})^{\T}$. It is of importance to note that $M_i$ is a function composed of at most first time derivatives as it is only composed of partial derivatives of $L$ and $\qdot$.
\end{subequations}
Moreover, if $K_{ij}$ is invertible, we can solve for $\qddot^i$ and get Newton's law \eqref{eq:N2}:
\begin{equation}
    \qddot^i =  (K^{-1})^{ij} M_j(\qdot^k, q^k,) = F(\qdot^k, q^k) \quad \text{or} \quad \mathbf{\qddot} = \textbf{K}^{-1}\textbf{M}(\dot{\textbf{q}}, \textbf{q}) =\mathbf{F}(\dot{\textbf{q}},\textbf{q}).
\end{equation}
If $K_{ij}$ is not invertible, the EoM does not take the regular Newtonian form, and means that the variables are not independent. Lagrangians for which $K_{ij}$ is not invertible are called degenerate and the invertibility is equivalent to what we will call the \textit{degeneracy condition}:
\begin{equation}\label{eq:degcond}
    \text{det}(K) = 0.
\end{equation}
This implies that $K_{ij}$ does not have full rank, i.e. at least one eigenvalue is equal to zero, and $K_{ij}$ has a non-trivial null-space. If $m=\mathcal{N}-\operatorname{rank} K$ are the number of null-eigenvectors $Y_{s}^{i}$, where $s=1, \ldots, m$ for which $Y_{s}^{i} K_{ij}=0$. Multiplying the Euler-Lagrange equations with the vector $Y_{s}^{i}$ from the left then reads 
\begin{equation}\label{eq:lagconst}
    0= Y^i_s  K_{ij}\qddot^j = Y_{s}^{i} M_i(\qdot^i, q^i,)=: \Phi_s(\qdot^i,q^i), \quad \text{or} \quad 0=\textbf{Y}^{T}_s \textbf{K}\ddot{\textbf{q}} = \textbf{Y}^{T}_s \textbf{M} = \Phi_s(\dot{\textbf{q}},\textbf{q}).
\end{equation}
We define constraint as the algebraic equation $\Phi_s=0$ only restricting the variables $q$ and $\qdot$. These are the variables lower than the dynamical equation of motion, which is a differential equation for $\qddot$. The $\Phi_s$ are thus the constraints arising from the EoM. We have therefore shown that 
\begin{equation}
    \text{det} (K) = 0 \;\;\; \Longleftrightarrow \;\;\; \text{existence of constraint},
\end{equation}
where a constraint is, as previously stated, an algebraic equation of lower-order variables than what is in the dynamical equation. For a first-order Lagrangian ,the constraint depends on $q$ and $\qdot$. Let us see what that means by an example.\\

% ----- ----- ----- EXAMPLE ---- ----- -----
\noindent\hrulefill \\

\noindent(E1) \textbf{Lagrangian constraints:} The following example is inspired by \citealp{Henneaux} and we will throughout the whole section go back to this example. Consider a system with the Lagrangian $L= \frac{1}{2}\left(\dot{q}^{1}-\dot{q}^{2}\right)^{2}-V(q^1,q^2)$. At first sight, there is no trivial reason why this Lagrangian might be problematic. Computing $K_{ij}$ explicitly gives
\begin{equation} \tag{E1.1}
    K = \left(\begin{array}{cc}
    \dfrac{\p^2 L}{\p ({\qdot^1})^2} &     \dfrac{\p^2 L}{\p {\qdot^1} \p {q^2}} \\
      \dfrac{\p^2 L}{\p {\qdot^1} \p {\qdot^2}} & \dfrac{\p^2 L}{\p ({\qdot^2})^2} \\
    \end{array}\right)
    = 
     \left(\begin{array}{cc}
    1 &     -1\\
      -1 & 1 \\
    \end{array}\right),
\end{equation}
which clearly shows that the determinant is equal to zero. As a result, we cannot solve to get linear independent EoM's for each individual $\qddot^i$.  What is the constraint then? Following \eqref{eq:lagconst}, we first compute $M_i$
\begin{subequations}\label{eq:langconstex}
\begin{align}
    M_i= \frac{\partial L}{\partial q^i} - \frac{\partial^2L}{\partial \qdot^i \partial q^j}\qdot^j &=
    \left(\begin{array}{c}
    \dfrac{\p L}{\p q^1} \\
       \dfrac{\p L}{\p q^2} \\
    \end{array}\right)
    -
    \left(\begin{array}{cc}
    \dfrac{\p^2 L}{\p \qdot^1 \p q^1} &   \dfrac{\p^2 L}{\p \qdot^1 \p q^2}\\
       \dfrac{\p^2 L}{\p \qdot^2 \p q^1} & \dfrac{\p^2 L}{\p \qdot^2 \p q^2} \\
    \end{array}\right) 
    \left(\begin{array}{c}
    \qdot^1\\
     \qdot^2 \\
    \end{array}\right) \tag{E1.2a} \\
    &= 
       \left(\begin{array}{cc}
    \dfrac{\p V}{\p q^1}   \\
       \dfrac{\p V}{\p q^2} \\
    \end{array}\right), \tag{E1.2b} 
\end{align}
\end{subequations}
where the second term is trivially zero since the potential term of the Lagrangian only depends on the coordinates and not the velocities. A null-eigenvector $Y^i$ to $K_{ij}$ is simply $(1,1)^{\T}$, and multiplying the above expression with $\textbf{Y}^{\T}$ from the left yields our constraint:
\begin{equation}\tag{E1.3}
    0 = 
    \left(\begin{array}{cc}
    1 & 1\\
    \end{array}\right)
    \left(\begin{array}{cc}
    \dfrac{\p V}{\p q^1}   \\
       \dfrac{\p V}{\p q^2} \\
    \end{array}\right)
    = 
    \dfrac{\p V}{\p q^1} + \dfrac{\p V}{\p q^2}   \\
    = 
    \Phi(q^1,q^2).
\end{equation}
Taking the constraint and the EoM into consideration (also called evaluating the system \textit{on-shell}) it is an algebraic expression resulting in a potential of the form $V(q^1-q^2)$, as otherwise, the derivatives would not cancel each other. Applying this to our Lagrangian would result in a Lagrangian only dependent on $q^1-q^2$.

Let us therefore now make a simple change of variables $\phi := q^1- q^2$ and $\psi=q^1+q^2$. This would give us the equivalent Lagrangian
\begin{equation} \tag{E1.4}
    L(\phidot, \phi, \dot{\psi}, \psi) = \frac{1}{2}\dot{\phi}^2 - V(\phi, \psi).
\end{equation}
The EoM for $\psi$ is
\begin{equation}
    \psi: \quad \pL{\psi} - \pL{\dot{\psi}} = \pdv{V}{\psi} = 0,
\end{equation}
an equation for $\psi$ to solve in terms of $\phi$. Consequently, we can write the potential function as $\widetilde{V}(\phi)=V(\phi,\psi(\phi))$ and thus end up with the final on-shell Lagrangian
\begin{equation}
    L(\phidot, \phi, \dot{\psi}, \psi)=\tfrac{1}{2}\phidot^2-\widetilde{V}(\phi).
\end{equation}
Something worth mentioning here is that when looking at the original system, its composition of two coordinates might make you think the velocity space (of all the $q^i$'s and $\qdot^i$'s) is of four dimensions. However, since that system is equivalent to the single variable system of $\phi$, the velocity space must be two dimensions. The explanation is that the Legendre transformation between the Lagrangian and Hamiltonian is not a bijection in the constrained two-variable system but will effectively reduce the dimension by two. On the other hand, the Lagrangian in the unconstrained single variable system will transform one-to-one to the Hamiltonian. The following sections will introduce the Hamiltonian perspective of this problem, giving us tools to understand it better.\\

\noindent\hrulefill

%%%%%%%%%%%%%%%% Primary and secondary constraints, and the Dirac algorithm %%%%%%%%%%%%%%%%%%%%% 

\subsection{Primary constraints}
In Hamiltonian formalism, we are interested in expressing the velocities in terms of the other phase variables ($q^i, p_i$), and momenta in particular, something that we often take for granted in the Legendre transformation of the Lagrangian:
\begin{equation}\label{eq:Hdef}
    H(q,p) =\dot{q}^{i} p_{i}(q, \dot{q})-L(q, \dot{q}).
\end{equation}
We will now look at how to treat a constrained Hamiltonian system, and how we can ensure that the Legendre transformation is a bijection. The transformation between Lagrangian and Hamiltonian formalism requires the mapping $\left(q^{i}, \dot{q}^{i}\right) \mapsto\left(q^{i}, p_{j}(q, \dot{q})\right)$, which in the unconstrained case is a one-to-one transformation of variables on the phase space. However, the existence of constraints will effectively reduce the phase space dimension and, thus, the bijective nature between the Hamiltonian and Lagrangian. In order to understand how this works, we begin looking at the canonical momenta. By definition,
\begin{equation}\label{eq:defmom}
 p_{i}\left(\dot{q}^{k}, q^{k} \right)=\frac{\partial L}{\partial \dot{q}^{i}}.
\end{equation}
Moreover, this allows us to express $K_{ij}$ as
\begin{equation}
    K_{ij} = \pdv{p_i}{\dot{q}^j}.
\end{equation}
This means that for a variation $\text{d}p^i$ we have that $dp_i = K_{ij}d\qdot^j$, and more importantly that $p_i$ are $\mathcal{N}$ independent variables if and only if $K$ is invertible such that $\qdot^{i}$ can be solved for coordinates and momenta. We can therefore conclude that invertibility of $K$ ensures us both a Newton-dynamical equation \eqref{eq:N2} and bijection between the velocities and momenta. 

Furthermore, we define the constraint surface $\psi_s=0$, in the phase space, restricting the possible momenta for the system, as a \textit{primary constraint}. A primary constraint is thus a condition of only the momenta, without using the equations of motion. These constraints may arise naturally dependent on the form of the Lagrangian, and before we continue we will study the previous example again, but now from a Hamiltonian perspective. \\

\noindent\hrulefill\\

% ------------- EXAMPLE ---------

\begin{figure}
    \centering
    \includegraphics[width=0.9\textwidth]{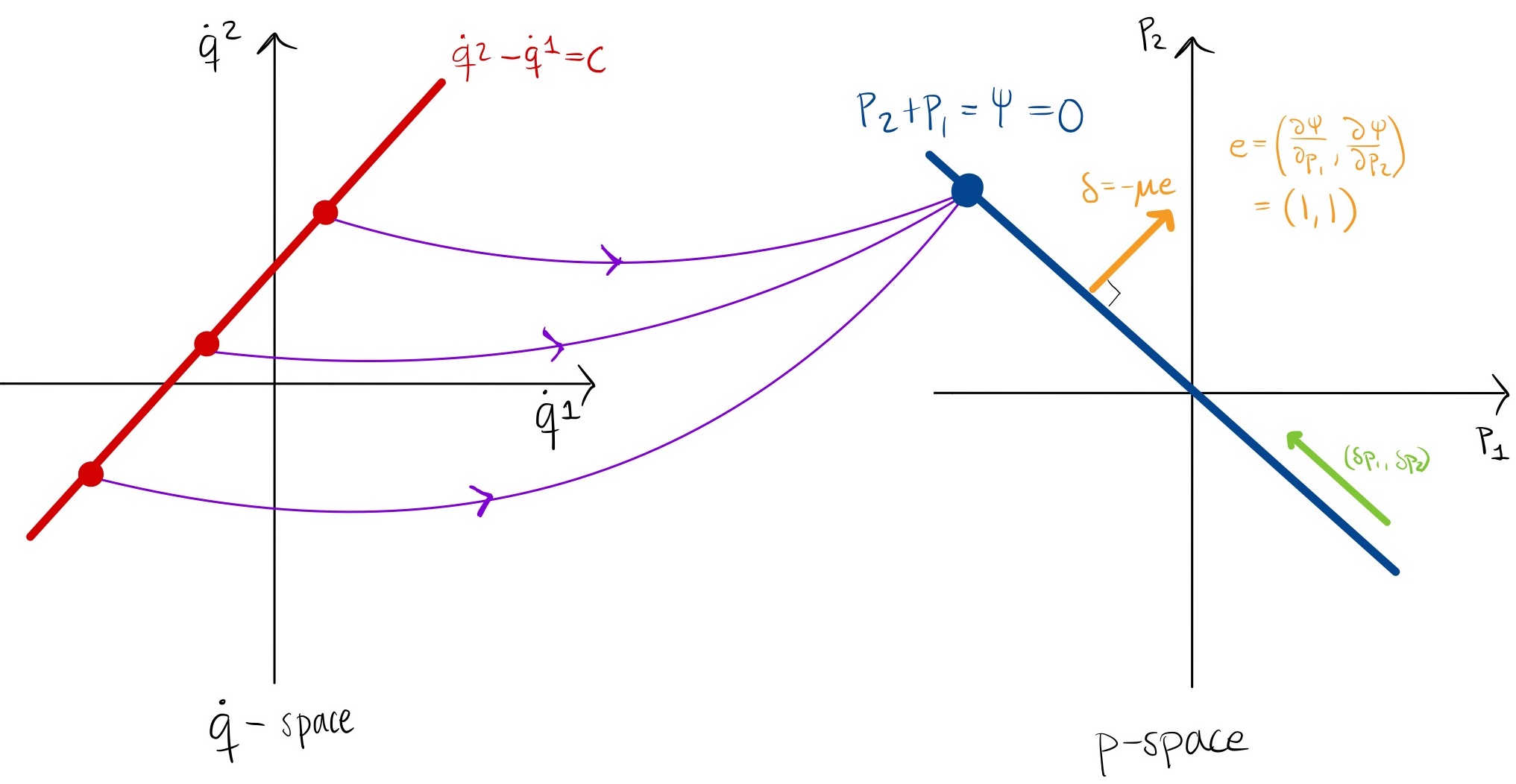}
    \caption{ [For example \citep{Henneaux}] An illustration of the transformation between velocities and momenta of the system $L= \tfrac{1}{2}\left(\dot{q}^{1}-\dot{q}^{2}\right)^{2}$. The blue line represents the constraint surface $p_1+p_2=0$ all $\qdot$-space is mapped on. Moreover, all $\dot{q}$ 's positioned on the red line  $\dot{q}^{2}-\dot{q}^{1}=c$  are mapped onto the same point $p_1=-c=p_2$. Moreover, the $\delta$ vector is presented as the orange vector, whereas a vector ($\delta p_1, \delta p_2$) is presented in green. For more details, see the main text}
    \label{fig:primconst}
\end{figure}

\noindent(E2) \textbf{Primary constraints:} Recall the system $L= \frac{1}{2}\left(\dot{q}^{1}-\dot{q}^{2}\right)^{2}-V(q^1,q^2)$. The canonical momenta are
\begin{equation} \tag{E2.1}
    p_{1}=\dot{q}^{1}-\dot{q}^{2}, \quad p_{2}=\dot{q}^{2}-\dot{q}^{1}, 
\end{equation}
and one can immediately spot a primary constraint $\psi=p_{1}+p_{2}=0$. We have already studied $K$ and arrived at the conclusion that its determinant is equal to zero and we thus have a degenerate system. We also saw that the constraint reduced the dimensions of phase space when transforming the Lagrangian. The Legendre transformation's building block is the transformation between the velocity and momenta, and we will now look at what that looks like. Figure \ref{fig:primconst} illustrates the mapping between $\qdot$-space and $p$-space. Note however that the constraint surface lays in full phase space $(q^1,q^2, p_1, p_2)$, but in this particular case we don't have a coordinate dependence on our primary constraint and only showing $p$-space is therefore sufficient. 

As we can see, all $\dot{q}$-space is mapped on the constraint surface, in this case a straight line, $p_{1}+p_{2}=0$ in $p$-space. Moreover, all the $\dot{q}$ 's positioned on the line $\dot{q}^{2}-\dot{q}^{1}=c$ are mapped on the same point $p_{1}=-c=-p_{2}$ on the constraint surface $\psi=0$. Consequently, the transformation $\dot{q} \rightarrow p$ is neither one-to-one nor onto. Because of this, the Hamiltonian is not defined in the whole phase space and we cannot use Legendre transformation at this stage. We will therefore now introduce necessary adjustments of our formalism, to ensure a correct Hamiltonian.

\noindent\hrulefill\\
\noindent
How can we guarantee that the Hamiltonian is always expressed in coordinates and momenta? A variation of the Legendre transformed Hamiltonian \eqref{eq:Hdef} reads
\begin{subequations}
\begin{align}
  \delta H(q^i,p_j) &= \dot{q}^{i} \delta p_{i}+p_{i} \delta \dot{q}^{i}-\frac{\partial L}{\partial \dot{q}^{i}} \delta \dot{q}^{i}-\frac{\partial L}{\partial q^{i}} \delta q^{i} \lref{\ref{eq:defmom}} \dot{q}^{i} \delta p_{i}-\frac{\partial L}{\partial q^{i}} \delta q^{i} \label{eq:varH1}\\
  &= \frac{\partial H}{\partial q^{i}} \delta q^{i}+\frac{\partial H}{\partial p_{i}} \delta p_{i} \label{eq:varH2},
\end{align}
where we have used the definition of momenta, assuming all information about constraints is included, and general function variation. From this we can see that $\delta H$ does only depend on variations of $q^i $ and $p_j$ and $H$ itself must therefore only depend on those variables.
\end{subequations}
Combining \eqref{eq:varH1} and \eqref{eq:varH2}, we get that $H$ must satisfy
\begin{equation}\label{eq:vecvarH}
    \left(\frac{\partial H}{\partial q^{i}}+\frac{\partial L}{\partial q^{i}}\right) \delta q^{i}+\left(\frac{\partial H}{\partial p_{i}}-\dot{q}^{i}\right) \delta p_{i}=0
    \Longleftrightarrow 
    \left(\frac{\partial H}{\partial q^{i}}+\frac{\partial L}{\partial q^{i}}, \frac{\partial H}{\partial p_{j}}-\dot{q}^{j}\right)\left(\begin{array}{l}
\delta q^{i} \\
\delta p_{j}
\end{array}\right)=0,
\end{equation}
where the vector $\left(\delta q^{i}, \delta p_{i}\right)$ is tangent to the primary constraint surface, that is the vector is tangent to every point $(q^i,p_i)$ laying on the constraint surface. Consequently,
\begin{equation}\label{eq:defdelta}
  \delta := \left(\frac{\partial H}{\partial q^{i}}+\frac{\partial L}{\partial q^{i}}, \frac{\partial H}{\partial p_{j}}-\dot{q}^{j}\right)
\end{equation}
must be normal to the constraint surface to satisfy \eqref{eq:vecvarH}. We can always find an orthonormal basis $\{\mathbf{e}_s\}$ for such surface by simply defining 
\begin{equation}
    \mathbf{e}_{s}:= \left(\frac{\partial \psi_{s}}{\partial q^{i}}, \frac{\partial \psi_{s}}{\partial p_{i}}\right),
\end{equation}
where $s = 1, \dots, m$ includes all the primary constraint functions $\psi_s$. We can now express all $\delta$ vectors using this basis, and for instance by setting $\delta=\sum_{s}- \mu^{s} e_{s}$ for some coefficients $\mu^{s}$, which might be functions on phase space. We can now rewrite \eqref{eq:defdelta} as
\begin{subequations}
\begin{align}
\frac{\partial H}{\partial q^{i}}+\frac{\partial L}{\partial q^{i}} &=-\mu^{s} \frac{\partial \psi_{s}}{\partial q^{i}}, \\
\frac{\partial H}{\partial p_{i}}-\dot{q}^{i} &=-\mu^{s} \frac{\partial \psi_{s}}{\partial p_{i}}.
\end{align}
\end{subequations}
Solving for $\qdot^i$ and $p_i$, and using the definition of momenta, we have now derived the Hamilton's equations:
\begin{subequations}
\begin{align}
\dot{q}^{i} &=\frac{\partial H}{\partial p_{i}}+\mu^{s} \frac{\partial \psi_{s}}{\partial p_{i}} = \frac{\partial\left(H+\mu^{s} \psi_{s}\right)}{\partial p_{i}}+\frac{\partial \mu^{s}}{\partial p_{i}} \psi_{s}, \\
\dot{p}_{i} &=-\frac{\partial H}{\partial q^{i}}-\lambda^{s} \frac{\partial \psi_{s}}{\partial q^{i}} = -\frac{\partial\left(H+\mu^{s} \psi_{s}\right)}{\partial q^{i}}+\frac{\partial \mu^{s}}{\partial q_{i}} \psi_{s}.
\end{align}
\end{subequations}
The rewritten terms, after the second equality signs, show that, up to terms that vanish on the primary constraint surface $\psi_s=0$, these equations can be seen as the usual Hamiltonian equations for the total Hamiltonian $H_T$ defined as
\begin{equation}\label{eq:defHtot}
    H_{T}:= H+\mu^{s} \psi_{s},
\end{equation}
for primary constraints $\psi_s=0$ on the constraint surface. Moreover, the Hamilton's equations can now be expressed more compact as
\begin{equation}
    \dot{q}^{i} \approx \{q^i,H_T\}, \quad \dot{p}_{i} \approx \{p_i,H_T\},
\end{equation}
where the \textit{weak equality} sign $\approx$ denotes an identity up to terms that vanish on the constraint surface. Additionally, the expression should not be set to zero before inserting it in Poisson brackets. Before we continue, let us consider the total Hamiltonian in our example.

\noindent\hrulefill\\

% ------------- EXAMPLE ---------

\noindent(E3) \textbf{Total Hamiltonian:} Now when we have the tools to create a correct Hamiltonian formalism, how will the Hamiltonian look for our system $L= \tfrac{1}{2}\left(\dot{q}^{1}-\dot{q}^{2}\right)^{2}-V(q^1,q^2)$? Recall the canonical momenta
\begin{equation} \tag{E3.1}
    p_{1}=\dot{q}^{1}-\dot{q}^{2}, \quad p_{2}=\dot{q}^{2}-\dot{q}^{1}, 
\end{equation}
and the primary constraint $\psi=p_{1}+p_{2}\approx0$. We eliminate $\qdot^1$ by using $\qdot^1=p_1+\qdot^2$. The total Hamiltonian by definition \eqref{eq:defHtot} is given by %(q^1,q^2, p_1, p_2)
\begin{align} 
    H_T &= p_1 \qdot^1 + p_2 \qdot^2 - \tfrac{1}{2}p_1^2 + V(q^1,q^2) +\mu\psi \tag{E3.2a}\\
    &= \frac{1}{2}p_1^2+ V(q^1,q^2) +(\qdot^2 + \mu)\psi \tag{E3.2b}
\end{align}
Furthermore, to visualise the vector $\delta$ defined in \eqref{eq:defdelta}, one can now look at the orange vector in Figure \ref{fig:primconst}, as $\delta$ is orthogonal to the constraint surface (in this case, the blue line). As described above, the vector $(\delta p_1, \delta p_2)$ is tangent to the constraint surface and is thus represented as the green vector in the figure.

\noindent\hrulefill\\

\subsection{Secondary constraints}
So far, we have shown the existence of primary constraints in Hamiltonian systems. However, in principle, we have only studied conditions of momenta at a given time, and we have not considered the time evolution of primary constraints. Requiring stability, e.g. that the constraints hold for all time, takes us to analyse the dynamics of primary constraints, and the Dirac algorithm in particular.

A primary constraint $\psi_s$ is preserved in time if $\dot{\psi}\approx0$. Evaluating the Poisson bracket $\{\psi_s, H_T\}$, and requiring it to be zero, yield
\begin{equation}\label{eq:secondaryconst}
        \{\psi_s, H_T\} =  \{\psi_s, H\} + \{\psi_s, \mu^s\}\psi_s +\{ \psi_s, \psi_t \} \mu^t \approx  \{\psi_s, H\} \{ \psi_s, \psi_t \} \mu^t  \approx 0,
\end{equation}
where $ \{\psi_s, \mu^s\}$ is not well-defined, however since it will always be multiplied by a constraint, the whole expression is weakly zero. Generally, $\{\psi_s, H_T\}$ is weakly non-zero (if it wasn't, the dynamics of $\psi$ would be trivial since then $\dot{\psi}\approx 0$ immediately), and instead generates a new constraint $\Pi_s$ simply defined as
\begin{equation}
      \Pi_s  := \{\psi_s, H_T\} \approx 0.
\end{equation}
 This is called a \textit{secondary} constraint, generated by the time evolution of a primary constraint. One can easily see how the time evolution of this constraint will generate a new one and so on, and this is the \textit{Dirac algorithm} \citep{Dirac}. By continuing this analysis and requiring stability in time for each constraint, you will gain a set of constraints, reducing degrees of freedom and ensuring a correct Hamiltonian formalism. To connect back to the Lagrangian constraints, the secondary constraint is related to the constraint $Y_s^iM_i$, whereas the primary constraint is related to the eigenvector $Y_s^i$.

How does the algorithm then end? In general, there are three cases to consider when requiring time stability like \eqref{eq:secondaryconst}. The first one is that the dynamics only depend on dynamical variables, and we get another constraint. Secondly, the dynamics is trivially $0$\footnote{This is generally an implication of a Gauge symmetry and that there are arbitrary functions left in the EoM. However, these functions do not affect the physical property of the system.} and lastly, that the only undefined variable in the expression is the coefficient $\mu^s$ and thus becomes an algebraic equation for them, which you can solve to fix the coefficients. This requires that the last term in \eqref{eq:secondaryconst} is non-zero.. The Dirac algorithm ends in the two last cases, whereas in the first case you need to keep requiring consistency of the constraints and will eventually end up in one of the last cases. Let's get back to our example again and apply this. \\

\noindent\hrulefill\\

% ------------- EXAMPLE ---------

\noindent(E4) \textbf{Secondary constraints:} Now considering the total Hamiltonian of the system, given by the previous example section, we have

\begin{equation} \tag{E4.1}
    H_T = \frac{1}{2}p_1^2+ V(q^1,q^2) +(\qdot^2 + \mu)\psi 
\end{equation}
We now want to require consistency of the primary constraint $\psi$. The time evolution of $\psi$ is given by 
\begin{align}
    \dot{\psi} = \{\psi, H_T\} &= \left\{ p_1 + p_2, \tfrac{1}{2}p_1^2 + V(q^1, q^2) + (\qdot^2 + \mu)\psi \right\} \tag{E4.2a}\\
    &= \poissonbracket{p_1}{V} + \poissonbracket{p_2}{V}+ \poissonbracket{p_1+p_2}{\qdot^2 + \mu}\psi,  \tag{E4.2b}
\end{align}
since $\poissonbracket{p_1}{p_2}=0$. Furthermore, we can evaluate the Poisson brackets for the potential term and the final expression is thus
\begin{align}
    \dot{\psi} &=  -\left( \pdv{V}{q^1}+ \pdv{V}{q^2}\right)+\poissonbracket{p_1+p_2}{\qdot^2 + \mu}\psi, \tag{E4.2b}\\
    & \approx-\left( \pdv{V}{q^1}+ \pdv{V}{q^2}\right) \approx 0. \tag{E4.2c}
\end{align}
Consequently, we get a secondary constraint that can be defined as
\begin{equation}\tag{E4.3}
    \Pi =  \left( \pdv{V}{q^1}+ \pdv{V}{q^2}\right)- \poissonbracket{p_1+p_2}{\qdot^2 + \mu}\psi \approx \left( \pdv{V}{q^1}+ \pdv{V}{q^2}\right)\approx 0
\end{equation}
Note that we again arrive at the same expression as for the Lagrangian constraint $Y_iM_i$ \eqref{eq:langconstex}. For the time evolution $\dot{\Pi}\approx0$ of $\Pi$, one again computes the Poisson bracket $\poissonbracket{\Pi}{H_T}$ and will arrive at an expression of the form
\begin{equation}\tag{E4.4}
    \dot{\Pi} = p_1\mathcal{A} + (\qdot^1+\mu)\mathcal{B} + \Psi \approx 0,
\end{equation}
where $\mathcal{A}$ and $\mathcal{A}$ are first-order terms and $\Psi$ is a term composed of terms multiplied by other constraints such that $\Psi \approx 0$. Generally, $\mathcal{A}$ and $\mathcal{B}$ are non-zero, which makes the above expression not a constraint but an equation for $\mu$. We can thus fix $\mu$ and the Dirac algorithm ends here.

\noindent\hrulefill\\

\noindent
In conclusion, we he have now seen 
\begin{equation}
    \text{det}(K)=0 \Longleftrightarrow \text{existence of primary constraint},
\end{equation}
which can then imply the existence of secondary constraints. We will in the next section use this knowledge to do a Hamiltonian analysis of the general second-order Lagrangian.

\section{Hamiltonian analysis of a general second-order Lagrangian}\label{sec:hamanalysis}
In this section, we will do Hamiltonian analyses of the second-order Lagrangian, both the single variable case and the case where such a system is coupled to a regular system. We will consider both degenerate and non-degenerate systems, using our knowledge from the previous section to evade the Ostrogradsky ghost. These analyses are inspired by the cases presented in \citealp{Ganz}, but are here done using full Hamiltonian formalism.

% ----- NON-DEGENERATE SINGLE VARIABLE ------ 
\subsection{Non-degenerate single variable}\label{sec:nondegsingle}
We will now consider the general non-degenerate Lagrangian $L(\phiddot, \phidot,\phi)$ briefly studied in section \ref{sec:existence}, but now from a Hamiltonian perspective.  Even if the initial Lagrangian is non-degenerate, our plan is to use Lagrange multipliers to practically work on degenerate regular systems. In other words, we want to use primary and secondary constraints (and hence degeneracy) in a regular system to analyse our initial exotic system. In general, the degeneracy will of course be different for a higher-derivatives context. Explicitly, $K_{ij}$ for a second-order system will now depend on the second-order derivative. Recall the second-order EL equation of $\phi$ as
\begin{equation}\label{eq:L2}
    \frac{\partial L}{\partial \phi}  -  \frac{\text{d}}{\text{d}t} \left\{\left(\frac{\partial L}{\partial \phidot} \right) + \dt \left(\frac{\partial L}{\partial \phiddot} \right) \right\}= 0,
\end{equation}
where the expression inside the curly brackets is the \textit{physical momenta} of this system. Adding Lagrange multiplier $\lambda$ we will for this problem consider an equivalent Lagrangian, with the action 
\begin{equation}
    S[\phi, q, \lambda] = \int \, \text{d}t \, \left( L(\qdot, q, \phi) + \lambda(\phidot-q)\right).
\end{equation}
Moreover,  the canonical momenta are
\begin{subequations}
\begin{align}
     &\phi: \quad P_\phi = \lambda\\
     &q:  \quad p = \pL{\qdot}(\qdot, q, \phi). \\
     &\lambda: \quad P_\lambda= 0 ,
\end{align}
\end{subequations}
and the only non-vanishing canonical Poisson brackets are $\poissonbracket{\phi}{P_\phi}=\poissonbracket{q}{p}=\poissonbracket{\lambda}{P_\lambda}=1$. From the definition of momenta, we  see two primary constraints $\Lambda := P_\lambda \approx 0$ and $\psi := P_\phi - \lambda \approx 0$. Before we enter the Hamiltonian formalism, we need to make sure that we can express the velocities in terms of the other phase space variables. One may ask if it is possible to invert $p=p(\qdot, q, \phi)$ to get $\qdot(p,q,\phi)$, and the answer is that this cannot be done analytically in general. However, according to the implicit function theorem, we can find such an expression locally, provided that we have non-degeneracy
\begin{equation}\label{eq:Hdeg}
    \frac{\partial^2 L}{\partial \qdot^2} = \frac{\p p}{\p \qdot} \neq 0
\end{equation}
to hold for the whole phase space, which in this case is true since $K_{ij}\neq 0$. We can therefore express $\qdot$ without involving $\lambda$, but most importantly in terms of momenta.
The total computed Hamiltonian is then
\begin{subequations}
\begin{align}\label{eq:H2}
    H_T &= P_\phi \phidot + p \dot{q}  + P_\lambda \dot{\lambda} -L(\phi, \dot{q}, q, \phi) - \lambda(\qdot-q) + \mu\Lambda + \nu\psi,\\
    &= P_\phi q + p\dot{q} + (\dot{\lambda} + \mu)\Lambda + (\phidot -q + \nu)\psi.
\end{align}
\end{subequations}
This expression is indeed linear in $P_\phi$. However, it is only problematic if the momenta is unconstrained. Let us now examine if this is the case.  Requiring consistency of the two primary constraints leads to $\dot{\Lambda}\approx 0$ and $\dot{\psi}\approx 0$. After some algebraic computation these yield
\begin{subequations}
\begin{align}
    \dot{\Lambda} &=  \dots \approx \phidot - q + \nu \approx 0 \\
    \dot{\psi} &= \dots \approx p\left\{P_\phi,\qdot \right\} - \left\{P_\phi,L \right\} -\dot{\lambda} - \mu \approx 0.
\end{align}
The Poisson brackets will not depend on $P_\phi$, as it will only depend on the coordinates. Consequently, we have now two equations for the coefficients $\mu, \nu$ and hence they do not generate new constraints but instead fix them. This is hence where the Dirac algorithm ends and we can conclude that $P_\phi$ is unconstrained and that the Hamiltonian is unbounded, as expected.
\end{subequations}

% ----- DEGENERATE SINGLE VARIABLE ------ 

\subsection{Degenerate single variable}
Moving on to consider a degenerate initial Lagrangian. The system is almost identical to the non-degenerate case in the previous section. However, now the degeneracy condition \eqref{eq:Hdeg} is equal to zero. This implies that the Lagrangian is at most linear in $\phiddot$. Accordingly, we can rewrite the Lagrangian in the most general form as
\begin{equation}
    \frac{\p^2L}{\p \phiddot^2}=0 \Longrightarrow L(\phiddot, \phidot,\phi)=\phiddot L_{0}(\phidot, \phi) + L_{2}(\phidot,\phi).
\end{equation}
If we define $L_1(\phidot, \phi):= \int \mathrm{d} \phidot L_0(\phidot, \phi)$, the action can then be rewritten, using integration by parts, as
\begin{equation}
  S[\phi] =\int \, \text{d}t \, \left(\phiddot L_{0}(\phidot, \phi) +L_{2}(\phidot, \phi)+\right) =  \int \, \text{d}t \,  \left( -\phidot\frac{\p L_1}{\p \phi}(\phidot,\phi)+L_2(\phidot,\phi) \right) + [\text{BT}].
\end{equation}
We have now reduced the initial degenerate Lagrangian to an equivalent Lagrangian with only one derivative, i.e a regular system. If we would have continued the Hamiltonian analysis we would have ended up with a primary constraint further generating a secondary constraint that would restrict $P_\phi$. Consistency of the secondary constraint would then fix the Lagrange multiplier. As a result, a degenerate single-variable Lagrangian thus does not suffer from an Ostrogradsky ghost. Therefore, it is of more interest to examine the more general case; degeneracy in an exotic system coupled to a regular system, which we will do in section \ref{sec:degcouple}. 

% ----- NON-DEGENERATE COUPLING ------

\subsection{Non-degenerate coupling}\label{sec:nondegcouple}
We are now going to look at the single variable system, coupled with the system of the regular variable $x$. Will coupling affect the existence of the Ostrogradsky ghost? As usual, we will use Lagrange multipliers to find an equivalent degenerate regular system. The analysis is similar to in section \ref{sec:nondegsingle}, but we are now having an additional variable $x$.

Accordingly, setting $q := \phidot$ and using the Lagrange multiplier $\lambda$, the joint action of the whole system is
\begin{align}
    S[q, \phi, \lambda; x] = \int \, \text{d}t \, \left( L(\qdot, q, \phi ; \dot{x}, x)+\lambda(\phidot-q) \right). 
\end{align}
The canonical momenta are
\begin{subequations}
\begin{align}
    P_\phi &= \lambda \\
    P_\lambda &= 0 \\ 
      p&=\frac{\partial L}{\partial \dot{q}}(\dot{q}, q, \phi ; \dot{x}, x)\\ p_x&=\frac{\partial L}{\partial \dot{x}}(\dot{q}, q, \phi ; \dot{x}, x),
\end{align}
and we again have two primary constraints $\Lambda := P_\lambda \approx 0$ and $\psi := P_\phi - \lambda \approx 0$.
\end{subequations}
Furthermore, since we have non-degeneracy we can express the velocities in terms of the momenta:
\begin{equation}
    \tfrac{\p^2L}{\p \qdot^2 } \left( =  \tfrac{\p^2L}{\p \phidot^2 }\right) = \tfrac{\p p}{\p \qdot}  \neq 0 \quad \text{and} \quad \tfrac{\p^2L}{\p \dot{x}^2 } = \tfrac{\p p_x}{\p \dot{x}}  \neq 0 \Longrightarrow \qdot=\qdot(p,p_x, q, \phi; x) \quad \text{and} \quad \dot{x}=\dot{x}(p,p_x, q, \phi; x). 
\end{equation}
We begin the Hamiltonian analysis by introducing the conjugate variables and their non-vanishing Poisson brackets as
\begin{equation}
     \{\phi, P_\phi\}=\{\lambda, P_\lambda\}=\{q, p\}=\{x, p_x\}=1.
\end{equation}
Using the invertibility of momenta and velocity, the total Hamiltonian is now given by
\begin{subequations}\label{eq:Hcouple}
\begin{align}
 H_T &= P_\phi \phidot + p \dot{q}+p_x\dot{x} +P_\lambda \dot{\lambda}-L(\qdot, q, \phi ;  \dot{x}, x) - \lambda(\phidot-q) +\mu\Lambda + \nu\psi\\
 &= P_\phi q + p_x \dot{x} + p \qdot -L(\qdot, q, \phi ;  \dot{x}, x)  + (\dot{\lambda}+\mu)\Lambda + (\phidot-q+\nu)\psi, \\
 &= H + P_\phi q,
 \end{align}
 where $H:=p_x \dot{x} + p \qdot -L(\qdot, q, \phi ;  \dot{x}, x)  + (\dot{\lambda}+\mu)\Lambda + (\phidot-q+\nu)\psi$.
\end{subequations}
The time evolution of the primary constraints is almost identical to the case without coupling, and we again end up with equations that fixes $\mu$ and $\nu$:
\begin{subequations}
\begin{align}
    \dot{\Lambda} &= \{\Lambda, H_T \}= \dots \approx \phidot - q + \nu \approx 0 \\
    \dot{\psi} &= \{\psi, H_T \} = \dots \approx \left\{\psi, H \right\} -\dot{\lambda} - \mu \approx 0,
\end{align}
where $H$ is defined above. 
\end{subequations}
Therefore, we can conclude that $P_\phi$ is linear in this case and the  Ostrogradsky ghost is present in the Hamiltonian.

\subsection{Degenerate coupling}\label{sec:degcouple}
As it turns out, coupling a regular variable to the exotic system was not enough to evade the Ostrogradsky ghost. On the other hand, the ghost was not present for the degenerate single-variable case. How about adding degeneracy to a coupled system? For $K\neq0$ we could naively try to use integration by parts again, however we would then arrive to an expression dependent on $\ddot{x}$. We therefore need to consider other approaches on how to treat this system.

Our starting point is from the previous section, with the action
\begin{align}
    S[q, \phi, \lambda; x] = \int \, \text{d}t \, \left( L(\qdot, q, \phi ; \dot{x}, x)+\lambda(\phidot-q) \right). 
\end{align}
and the primary constraints $\Lambda := P_\lambda \approx 0$ and $\psi := P_\phi - \lambda \approx 0$. Continuing the analysis in the previous section, we now require that the degeneracy condition \eqref{eq:degcond} is fulfilled for the total Lagrangian of the coupled system. We expect the introduced degeneracy to generate an additional primary constraint. In matrix form, $\mathbf{K}$ for this system reads
\begin{equation}
  \left(\begin{array}{c}
\delta p \\
\delta y
\end{array}\right)
= \mathbf{K}
  \left(\begin{array}{c}
 \delta \qdot \\
\delta \dot{x}
\end{array}\right)
=   \left(\begin{array}{cc}
\dfrac{\p^2 L}{\p \qdot^2} & \dfrac{\p^2 L }{\p \dot{x} \p \qdot}  \\
\dfrac{\p^2 L }{\p \dot{x} \p \qdot}  & \dfrac{\partial L}{\p^2 \dot{x}^2} 
\end{array}\right)
  \left(\begin{array}{c}
 \delta \qdot \\
\delta \dot{x}
\end{array}\right)
=
  \left(\begin{array}{cc}
\dfrac{\partial p}{\partial \qdot} & \dfrac{\partial p}{\partial \dot{x}}  \\
 \dfrac{\partial p}{\partial \dot{x}} & \dfrac{\partial p_x}{\partial \dot{x}} 
\end{array}\right)
  \left(\begin{array}{c}
 \delta \qdot \\
\delta \dot{x}
\end{array}\right).
\end{equation}
Note that we chose to write $\frac{\partial p}{\partial \dot{x}}$ instead of  $\frac{\partial p_x}{\partial \qdot} $, to show the symmetric structure of $\mathbf{K}$, since they are equal for a $\mathcal{C}^2$ function.
The degeneracy condition is now
\begin{equation}
   \text{det } \mathbf{K}=0 \Longleftrightarrow \left(\dfrac{\partial p}{\partial \qdot}\right) \left( \dfrac{\partial p_x}{\partial \dot{x}}\right)  - \left( \dfrac{\partial p}{\partial \dot{x}}\right)^2 = 0.
\end{equation}
We will now explore how this equation will be satisfied for our system.

\subsubsection{Case 0: All terms are non-zero}
Assume that we have
\begin{equation}
    \dfrac{\partial p}{\partial \dot{x}} \neq 0 \quad \text{and} \quad  \dfrac{\partial p_x}{\partial \qdot}\neq 0.
\end{equation}
Then we must have  
\begin{equation}
   \frac{\p p}{\p \dot{x}} \frac{\p p_x}{\p \qdot} = \frac{\p p}{\p \qdot} \frac{\p p_x}{\p \dot{x}}. 
\end{equation}
Since the left-hand side is non-zero, so must the right-hand side. As a result, we in principle could invert (locally) both of the velocities in terms of the momenta:
\begin{subequations}
\begin{align}
     \frac{\p p}{\p \qdot} \neq 0 \Longrightarrow p(\qdot,...) &\xRightarrow{\text{locally}} \qdot(p,..),\\
    \frac{\p p_x}{\p \dot{x}} \neq 0 \Longrightarrow p_x(\dot{x},...) &\xRightarrow{\text{locally}} \dot{x}(p_x,..).
\end{align}
However, doing this for both velocities simultaneously contradicts the degeneracy condition as then det$(K)\neq0$. Therefore, it is of more interest to look at the cases where not all terms are non-zero. The first one is when all of the terms are zero, and the others are when one of the factors in the first term is non-zero.
\end{subequations}

\subsubsection{Case 1: K=0}
The first trivial case is when 
\begin{equation}
     \dfrac{\partial p}{\partial \qdot} = \dfrac{\partial p_x}{\partial \dot{x}} =  \dfrac{\partial p}{\partial \dot{x}} =0
\end{equation}
which naturally leads to $K$ being identical zero. As a result, this must mean that the momenta do not depend on the velocities at all. Or explicitly, 
\begin{equation}
    p = P(q, \phi; x)  \quad \text{and} \quad p_x = X(q, \phi; x), 
\end{equation}
where $P:=\frac{\p L}{\p \qdot}$ and $X:=\frac{\p L}{\p \dot{x}}$ are functions in phase space. Since we cannot solve these expressions for the velocities in terms of momenta, we get \textit{two} primary constraints:
\begin{subequations}\label{eq:twoconst}
\begin{align}
    \chi_p= p - P(q, x, \phi) &\approx 0 \\
    \chi_x = p_x - X(q, x, \phi) &\approx 0.
\end{align}
\end{subequations}
The analysis of this case is simpler than the following cases as it is very similar to what happened to the single variable case. One does not need to enter Hamiltonian formalism to examine whether there is an Ostrogradsky ghost or not. Let us therefore sort out what the terms being zero means for the Lagrangian. First, we have
\begin{equation}
   \frac{\p^2 L}{\p \qdot^2}=0 \Longrightarrow L = \qdot L_0 (q, \phi; \dot{x}, x) + L_2 (q, \phi; \dot{x}, x) \sim \phiddot L_0 (\phidot, \phi; \dot{x}, x) + L_2 (\phidot, \phi; \dot{x}, x),
\end{equation}
where we have used $\sim$ to show that they are identical up to a Lagrange multiplier-term $\lambda(\phidot - q)$. 
At this stage, it is not possible to reduce the order of the Lagrangian. However, now imposing $\frac{\p^2 L}{\p \dot{x}\p \qdot }=0$ implies that $L_0$ does not depend on $\dot{x}$:
\begin{equation}
   \frac{\p^2 L}{\p \dot{x}\p \qdot }=0 \Longrightarrow L  \sim \phiddot L_0 (\phidot, \phi; x) + L_2(\phidot, \phi; \dot{x}, x) .
\end{equation}
Note that starting with $\frac{\p^2 L}{\p \dot{x}^2}=0$ would lead to the same results. If we now define $L_1 := \int \text{d}\phidot \, L_0(\phidot, \phi; x)$, we can reduce the Lagrangian to one of first order using integration by parts:
\begin{equation}
    L \sim  L_2(\phidot, \phi; \dot{x}, x) - \frac{\p L_1}{\p \phi} \phidot - \frac{\p L_1}{\p x} \dot{x} + [\text{BT}].
\end{equation}
The conclusion here is therefore identical to the degenerate single-variable case; there is no Ostrogradsky ghost as a Hamiltonian analysis will end in constraints that fixes the Lagrange multiplier and restrict the momenta $P_\phi$. What made the reduction possible was the fact that both $\frac{\p^2 L }{\p \dot{x} \p \qdot}$ and $\frac{\p^2 L }{\p\qdot^2}$ are zero. For the following cases, this is not true and we thus need other tools to examine the Ostrogradsky ghost.

\subsubsection{Case 2 \& 3: One of the first terms is non-zero}
We are now going to assume that
\begin{equation}
    \dfrac{\partial p_x}{\partial \dot{x}} \neq 0.
\end{equation}
We will see that this implies that the other terms are zero. Our goal is, as usual, to find a way to express the velocities in terms of the phase space variables, in particular the momenta. Now, the above expression implies that we can locally express $\dot{x}$ in terms of $p_x$: $\dot{x}=\dot{x}(p_x, \qdot; q, x, \phi)$. Consequently, $p$ can then be locally expressed in terms of $p_x$ since
\begin{equation}
    p=p(\dot{q}, \dot{x}(p_x, \qdot; q, x, \phi) ; q, x, \phi)= p(\dot{q}, p_x ; q, x, \phi).
\end{equation}
Does $p$ depend on $\qdot$? According to the degeneracy condition, we now have
 \begin{equation}
    \left( \frac{\p p}{\p \qdot} \right)=\left(\frac{\p p}{\p \dot{x}}\right)^2 \left(\frac{\p p_x}{\p \dot{x}}\right)^{-1}.
 \end{equation}
 If $\frac{\p p}{\p \qdot} \neq 0$, we are back to Case $0$ and can again express both velocities in terms of the momenta simultaneously, and it contradicts the degeneracy condition. We therefore conclude that $p$ cannot depend on $\qdot$ and we finally get a primary constraint
\begin{subequations}
\begin{equation}\label{eq:const}
    \chi_p:= p - \mathcal{P}(p_x ; q, x, \phi) \approx 0,
\end{equation}
where 
\begin{equation}
    \mathcal{P} := \frac{\p L}{\p \qdot}\bigg|_{\dot{x}=\dot{x}(p_x, q, x, \phi)},
\end{equation}
is a function in phase space.

Using the same analogy, for the case when $\frac{\partial p}{\partial \qdot} \neq 0$, we arrive to a primary constraint of $p_x$
\begin{equation}
     \chi_x := p_x - \mathcal{X}(p ; q, x, \phi) \approx 0,
\end{equation}
\end{subequations}
for a function $\mathcal{X}:=  \dfrac{\p L}{\p \dot{x}}\bigg|_{\qdot=\qdot(p, q, x, \phi)}$ in phase space.

So far, we have shown that degeneracy implies at least one primary constraint of the system's momenta. Now we will continue the analysis and show that this further implies the existence of secondary constraints, for which they can eliminate a ghost degree of freedom.

Now, the analysis will look similar to both of these cases, so wee chose case $2$. We then have $\p p_x/ \p \dot{x} \neq 0$ \eqref{eq:const}, and the primary constraint $\chi_p := p - \mathcal{P}(p_x ; q, x, \phi) \approx 0$. This primary constraint will add up to the the other primary constraint we had in the non-degenerate case. The total Hamiltonian $H_T$ is then given by
\begin{subequations}
\begin{align}
    H_T&= P_\phi \phidot + p \dot{q}+p_x\dot{x} +P_\lambda \dot{\lambda}-L(\qdot, q, \phi ;  \dot{x}, x) - \lambda(\phidot-q) +\mu\Lambda + \nu\psi + \sigma \chi_p,\\
    &= P_\phi q + p_x \dot{x} + p \qdot -L(\qdot, q, \phi ;  \dot{x}, x)  + (\dot{\lambda}+\mu)\Lambda + (\phidot-q+\nu)\psi + \sigma \chi_p, \\
    &= H+P_\phi q+\sigma \chi_p,
\end{align}
\end{subequations}
where $H=p_x \dot{x} + p \qdot -L(\qdot, q, \phi ;  \dot{x}, x)  + (\dot{\lambda}+\mu)\Lambda + (\phidot-q+\nu)\psi$ was defined in \eqref{eq:Hcouple}, and $\sigma$ is as usual a function in phase space. Consistency of $\Lambda$ and $\psi$ is almost identically to the previous section:
\begin{subequations}
\begin{align}
    \dot{\Lambda} &= \{\Lambda, H_T \} = \dots \approx \phidot - q + \nu \approx 0, \\
    \dot{\psi} &= \{\psi, H_T \} = \dots \approx \left\{\psi, H \right\} -\dot{\lambda} - \mu - \sigma\{P_\phi, \mathcal{P}\}\approx 0,
\end{align}
where the first equation fixes $\nu$ and the other fixes $\mu$ in terms of $\sigma$.
\end{subequations}
The time evolution of our new primary constraint is explicitly computed by the Poisson bracket:
\begin{subequations}
\begin{align}
\dot{\chi_p} =\left\{\chi_p, H_{T}\right\} 
& =\left\{p-\mathcal{P}, H +P_\phi q\right\} + \{\chi_p, \sigma\}\chi_p \\
&=-P_\phi + \theta(p, p_x ; q, x, \phi; \dot{\lambda}; \mu, \nu) + \{\chi_p, \sigma\}\chi_p, \\
&\approx \theta(p, p_x ; q, x, \phi; \dot{\lambda}; \mu, \nu) - P_\phi, \\
&\approx \theta(p, p_x ; q, x, \phi; \dot{\lambda}) - P_\phi \big|_{\dot{\Lambda}\approx\dot{\psi}\approx0}, \\
&\approx 0 \notag
\end{align}
\end{subequations}
 where $\theta:=\left\{p-\mathcal{P}, H\right\}$. As a result, after imposing consistency of $(\mu,\nu)$  we obtain 
\begin{equation}\label{eq:lambdaconst}
    \Theta := P_\phi-\theta(p, p_x ; q, x, \phi; \dot{\lambda})- \{\chi_p, \sigma\}\chi_p \approx 0.
\end{equation}
This expression does not depend on any multipliers and is thus a secondary constraint that can be solved for $P_\phi$ in terms of the other phase space variables. Specifically, $P_\phi$ cannot take arbitrary values anymore and there is no longer an apparent reason to why the Hamiltonian would be unbounded from below.

To finish the Hamiltonian analysis, we will now look at the stability of $\Theta$ under time evolution;
\begin{subequations}
\begin{align}
    \dot{\Theta} = \{\Theta, H_T \} &= \dots \approx \{\Theta, H \} + P_\phi\{\Theta, q\} + \{\Theta, P_\phi\}q  + \sigma\{\Theta, \chi_p\}  \approx 0.
\end{align}
\end{subequations}
Since $\{\Theta, \chi_p\}$ generally is non-zero\footnote{The interested reader can look up first and second class constraints. This statement implies that the primary and secondary constraints are generally of second class. See for instance \citealp{Dirac,Henneaux}.}, this equation is an algebraic expression for which we can solve for and fix $\sigma$, and does not generate a new constraint. Hence, this is where the Dirac algorithm ends.

Finally, for the first case of the degeneracy condition, we end up with three primary constraints:
\begin{subequations}
\begin{align}
     \Lambda &= P_\lambda \approx 0, \\
     \psi &= P_\phi - \lambda \approx 0, \\
    \chi_p &= p - P(p_x ; q, x, \phi) \approx 0,
\end{align}
\end{subequations}
and four secondary constraints:
\begin{subequations}
\begin{align}
  \dot{\Lambda} & \approx \phidot - q + \nu \approx 0, \\
    \dot{\psi} &\approx \left\{\psi, H \right\} -\dot{\lambda} - \mu - \sigma\{P_\phi, P\}\approx 0, \\
    \Theta &\approx P_\phi-\theta(p, p_x ; q, x, \phi; \dot{\lambda}) \approx 0, \\
    \dot{\Theta} & \approx \{\Theta, H \} + P_\phi\{\Theta, q\} + \{\Theta, P_\phi\}q  + \sigma\{\Theta, \chi_p\}  \approx 0.
\end{align}
\end{subequations}

The secondary constraints $\dot{\Lambda} \approx \dot{\psi} \approx \dot{\Theta}\approx 0$ impose that we fix the three coefficients $(\mu, \nu, \sigma)$ whereas $\Theta$ finally restricted the physical momenta $P_\phi$. As a result, we successfully eliminated the Ostrogradsky ghost and there is no apparent reason to why the Hamiltonian would be unbounded. 

As previously stated, the third case will provide a similar analysis, and Case $1$ of $K=0$ also eliminated the ghost. We can therefore conclude that the Ostrogradsky ghost is eliminated in all of the three cases where the degeneracy condition is satisfied.

\section{Generalised Hamiltonian analysis}\label{sec:genHA}
Conducting Hamiltonian analysis for several cases of the second-order Lagrangian, one could perhaps see the potential of generalise the analysis. Indeed, this section aims to present an heuristic motivation to the generalisation of the Hamiltonian constraint analysis of primary and secondary constraints, to a system of higher derivatives.

We begin by considering a Lagrangian $L(q^{(n)}, q^{(n-1)}, \dots, q)$, where the parentheses denote the order of time derivative, e.g. $q^{(1)}=\qdot$. The equivalent first-order system would be of $n+1$ variables, which we can define as
\begin{equation}
    x_k:= q^{(k)}, \quad  k=0,1,\dots,n-1.
\end{equation}
We have now obtained a set of variables $x_0=q, x_1=\qdot, x_2=\qddot, \dots, x_{n-1}= q^{(n-1)}, \dot{x}_{n-1}= q^{(n)} $. Substituting this into the initial Lagrangian yields $L(\dot{x}_{n-1}, x_{n-1}, \dots, x_1, x_0)$. However, the equivalent Lagrangian is, as usual, given by using Lagrange multipliers $\lambda^k$:
\begin{subequations}
\begin{align}
 \widetilde{L} & = L(\dot{x}_{n-1}, x_{n-1}, \dots, x_1, x_0)+\lambda^0 (\dot{x}_0 -x_1 ) + \lambda^1(\dot{x}_1- x_2 ) +\dots + \lambda^{n-2}( \dot{x}_{n-2} -x_{n-1} ) \\
 &=  L(\dot{x}_{n-1}, x_{n-1}, \dots, x_1, x_0) + \sum_{k=0}^{n-2} \lambda^k(\dot{x}_k -x_{k+1}),
\end{align}
where we have reduced the initial $n$-order system to one of at most first-order.
\end{subequations}
The canonical momenta are now simply
\begin{subequations}
\begin{align}\label{eq:defpk}
    p_k := \frac{\p \widetilde{L}}{\p \dot{x_k}} =
    \begin{cases}
    \lambda^k, \quad &\text{for } k=0,1,\dots, n-2 \\
    \dfrac{\p L}{\p \dot{x}_{n-1}}, \quad &\text{for } k=n-1.
    \end{cases}
\end{align}
and 
\begin{equation}
 p_{\lambda^k} := \frac{\p \widetilde{L}}{\p \dot{\lambda}^k} = 0, \quad \text{for } k=0,1,\dots, n-2.
\end{equation}
\end{subequations}
We can observe $2(n-1)$ numbers of primary constraints $\psi_s:= p_s -\lambda^s\approx 0$ and $\pi_s:= p_{\lambda^s}\approx 0$, where $s=0,1,\dots, n-2$. Now we have everything we need to compute the total Hamiltonian:
\begin{equation}
    H_T = \sum_{k=0}^{n-1} p_k\dot{x}_k +P_{\lambda^k}\dot{\lambda}^k- L(\dot{x}_{n-1}, x_{n-1}, \dots, x_1, x_0) +  \sum_{k=0}^{n-2} \left[ \lambda^k(\dot{x}_k -x_{k+1}) + \mu^k \psi_k + \nu^k\pi_k \right],
\end{equation}
where $\mu^k$ and $\nu^k$ are functions on phase space. From this, one must require consistency of the primary constraints and continue until the Dirac algorithm ends. Indeed, a Hamiltonian analysis of this system will require a lot of Dirac machinery, although it presumably provides interesting results.

One can expect for a non-degenerate Lagrangian that  $\dot{\psi}_s \approx0$ och $\dot{\pi}_s\approx0$ will be equations for  $\mu_s$ och $\nu_s$,  and thus there will be no secondary constraints to evade a Ostrogradsky ghost.

On the other hand, for the degenerate case, we expect that the $p_{n-1}$, defined in equation \eqref{eq:defpk} will lead to a primary constraint and then generate secondary constraints which will eliminate a ghost. However, $H_T$ will still be linear in other canonical momenta, something that can take both negative and positive values, and as a result, the total Hamiltonian will be unbounded from below. Therefore, one will need more secondary constraints to eliminate all ghosts and avoid an Ostrogradsky instability.

Regardless, conducting a Hamiltonian analysis of a general-order Lagrangian and making explicit calculations would be an interesting topic for future study but out of scope (and margin) for this thesis.

\section{Discussion and conclusions}\label{sec:cd}
In this thesis, we began by showing the existence of the Ostrogradsky ghost in the second-order Lagrangian by presenting Hamiltonians with negative kinetic energy and linearity in unrestricted physical momenta. Furthermore, we have studied the essentials of Hamiltonian formalism to carry out Hamiltonian analyses of constrained second-order systems. Additionally, we presented an outline of a future generalised Hamiltonian analysis of higher-order systems.

Ultimately, all non-degenerate second-order Lagrangians considered contained an Ostrogradsky ghost. On the other hand, if the Lagrangian is degenerate, we conclude that the ghost can be evaded. The degeneracy will either reduce the higher-order Lagrangian to one that is first-order or generate secondary constraints that restrict the physical momenta in the Hamiltonian. Therefore, higher-order degenerate systems can be healthy by correctly using this thesis's methods, focusing mainly on the total Hamiltonian.

However, the Hamiltonian analysis becomes increasingly complicated as multiple variables are considered. A Lagrangian approach might be more convenient than the Hamiltonian formalism of higher-order systems, as we could avoid the mechanisms of Dirac's algorithm. Therefore, future work should include a more in-depth analysis of Lagrangian formalism, perhaps starting from the definition of the energy function. Moreover, the absence of an Ostrogradsky ghost does not necessarily mean the Hamiltonian is no longer unbounded. It could still be, but for reasons out of the scope of this thesis.
In essence, there is a lot to learn about higher-order derivative theories and, most importantly, that they can be correctly used to describe the physical properties of a system. Questioning what we are used to might sometimes lead to new insights into the physics we know, and on that note, this thesis will end with a memorable quote from Isaac Newton:

\begin{center}
    \textit{No great discovery was ever made without a bold guess.}\\
    - Isaac Newton
\end{center}

% Finally, we have discussed the general-order energy function from a Lagrangian perspective. We can conclude that the non-degenerate Lagrangian of order $n>1$ suffers from the Ostrogradsky instability, as it always produces a Hamiltonian linear in the unconstrained physical momenta. 

% As we saw in section \ref{sec:efunc}, the proof of the Ostrogradsky theorem in the Lagrangian formalism has many benefits over the Hamiltonian formalism, as we could avoid the machinery of Diracs algorithm. 

% One could however expect the following results:
% \begin{itemize}
%     \item[]\textbf{Trivial dynamics of the primary constraints:} All $\dot{\psi_k}\approx 0$, and we get only equations that fix $\mu^k$ and no secondary constraint.
%     \item[]\textbf{Degenerate}: Having $\frac{\p^2 L}{\p (q^{(n)})^2}=0$ would lead to a secondary constraint restricting the momenta $p_{n-2}$. However, $H_T$ would then still be linear in $p_{n-3}, p_{n-4}, \dots$, and additional secondary constraints would thus be needed to elevate all the ghosts.
%     %according to similar arguments as in section \ref{sec:efunc}.
% \end{itemize}

\section*{Acknowledgements}
The author wants to thank Joakim Flinckman for useful discussions, teaching and overall support. In particular, his advice that some things can be \textit{two coins of the same side}.
\bibliographystyle{mnras}
\bibliography{disk}

%%%%%%%%%%%%%%%%% APPENDICES %%%%%%%%%%%%%%%%%%%%%
\appendix

%\section{Useful formulas and relations}\label{sec:appendix}
\section{Second-order Euler-Lagrange equation}\label{sec:EL}
We will derive the EL equations by the classical method of considering it as the solution to an optimisation problem. Consider a Lagrangian $L(\phiddot^i, \phidot^i , \phi^i,t)$,
\begin{equation}
    S\left[\phi \right]=\int_{t_1}^{t_2} \, \text{d}t \, L\left(\phiddot^i, \phidot^i , \phi^i,t\right).
\end{equation}
The equation of motions are the paths following the principle of least action, leading us to ask the question of how we can minimise a variation $\delta S$. Consider a small variation of the path:
\begin{equation}
    \phi(t) \rightarrow \phi(t) + \delta \phi(t),
\end{equation}
where consistency with the boundary value problem requires that the variation $\delta \phi(t)$ and its derivatives vanish at the end points $t_1$ and $t_2$ of the integral. That is, $\delta \phi(t_1)=\delta \phi(t_2)=\delta \phidot(t_1)=\delta \phidot (t_2) =0$. Now, a variation in the action is given by
\begin{subequations}
\begin{align}
\delta S[\phi^i]&= \int_{t_1}^{t_2}\, \text{d}t \, \delta  L\left(\phiddot^i, \phidot^i , \phi^i,t\right), \\
&= \int_{t_1}^{t_2}\, \text{d}t \,\left(\frac{\partial L}{\partial \phi^i } \delta \phi^i+\frac{\partial L}{\partial \phidot^i} \delta \phidot^i+\frac{\partial L}{\partial \phiddot^i} \delta \phiddot^i\right). 
\end{align}
We can now use integration by parts to reduce the second-order derivative $\delta \phiddot^i$:
\begin{align}
\delta S[\phi^i]&= \int_{t_1}^{t_2} \, \text{d}t \, \left(\frac{\partial L}{\partial \phi^i} \delta \phi^i+\frac{\partial L}{\partial \phidot^i} \delta \phidot^i-\dt \frac{\partial L}{\partial \phiddot^i} \delta \phidot^i\right) +\left[\frac{\partial L}{\partial \phiddot^i} \delta \phidot^i\right]_{t_1}^{t_2}, \\
&= \int_{t_1}^{t_2} \, \text{d}t \, \left(\frac{\partial L}{\partial \phi^i} \delta \phi^i+\left(\frac{\partial L}{\partial \phidot^i}-\dt \frac{\partial L}{\partial \phiddot^i}\right) \delta \phidot^i\right) +\left[\frac{\partial L}{\partial \phiddot^i} \delta \phidot^i\right]_{t_1}^{t_2}.
\end{align}
Again, using integration by parts on $\delta \phidot^i$:
\begin{align}
\delta S[\phi^i]&= \int_{t_1}^{t_2}\, \text{d}t \, \left(\frac{\partial L}{\partial \phi^i} \delta \phi^i-\left(\dt \frac{\partial L}{\partial \phidot^i}-\ddtt \frac{\partial L}{\partial \phidot^i}\right) \delta \phi^i\right), \\
& \quad+\left[\left(\frac{\partial L}{\partial \phidot^i}-\dt \frac{\partial L}{\partial \phidot^i}\right) \delta \phi^i\right]_{t_1}^{t_2}+\left[\frac{\partial L}{\partial \phiddot^i} \delta \phidot^i\right]_{t_1}^{t_2}, \\
&= \int_{t_1}^{t_2} \, \text{d}t \, \left(\left\{\frac{\partial L}{\partial \phi^i}-\dt \frac{\partial L}{\partial \phidot^i}+\ddtt \frac{\partial L}{\partial \phidot^i}\right\} \delta \phi^i \right), \\
 & \quad +\left[\left(\frac{\partial L}{\partial \phidot^i}-\dt \frac{\partial L}{\partial \phidot^i}\right) \delta \phi^i\right]_{t_1}^{t_2}+\left[\frac{\partial L}{\partial \phidot^i} \delta \phiddot^i\right]_{t_1}^{t_2}\\
 &= \int_{t_1}^{t_2} \, \text{d}t \, \left(\left\{\frac{\partial L}{\partial \phi^i}-\dt \frac{\partial L}{\partial \phidot^i}+\ddtt \frac{\partial L}{\partial \phidot^i}\right\} \delta \phi^i \right),
\end{align}
\end{subequations}
where the square brackets vanish as the end points are fixed according to the boundary conditions of $\delta \phiddot^i$ and $\delta \phi$. Now, principle of least action requires $\delta S =0$, which happens when $L$ satisfies the equation of the expression inside the curly brackets being zero:
\begin{equation}
    \frac{\partial L}{\partial \phi^i}- \dt \left(\frac{\partial L}{\partial \phidot^i}\right)+\ddtt\left(\frac{\partial L}{\partial \phiddot^i}\right)=0,
\end{equation}
which is exactly the Euler-Lagrange equation for this system.

Moreover, with not much more effort, one can generalise the above arguments to a Lagrangian $L(\phi^{\tiny i(n)}, \phi^{\tiny i (n-1)} \dots, \phidot^i , \phi^i)$, where $(k)$ denotes the order of the time derivative. We needed to use integration by part twice for the second-order system, and one expects to need to integrate by parts $n$ times for a $n$-order Lagrangian. As a result, we will get alternating terms and the final expression of the general Euler-Lagrange equation is thus
\begin{equation}\label{eq:EL}
    \sum_{k=0}^{n}(-1)^k \frac{\mathrm{d}^k}{\mathrm{d}t^k}\left( \pL{\phi^{\tiny i(k)}}\right) = 0.
\end{equation}

\end{document}